# Toplayer-Dependent Crystallographic Orientation Imaging in the Bilayer Two-Dimensional Materials with Transverse Shear Microscopy

Sabir Hussain,[1,2,4,#] Rui Xu,[1,#] Kunqi Xu,[3] Le Lei,[1] Shuya Xing,[1] Jianfeng Guo,[1] Haoyu Dong,[1] Adeel Liaqat,[2,4] Rashid Iqbal,[2,4] Muhammad Ahsan Iqbal,[2,4] Shangzhi Gu,[1] Feiyue Cao,[1] Yan Jun Li,[5] Yasuhiro Sugawara,[5] Fei Pang,[1] Wei Ji,[1] Liming Xie,[2,4] Shanshan Chen,[1,*] Zhihai Cheng[1,*]

[1]Beijing Key Laboratory of Optoelectronic Functional Materials & Micro-nano Devices, Department of Physics, Renmin University of China, Beijing 100872, China

[2]CAS Key Laboratory of Standardization and Measurement for Nanotechnology, CAS Center for Excellence in Nanoscience, National Center for Nanoscience and Technology, Beijing 100190, China

[3]Key Laboratory of Inorganic Functional Materials and Devices, Shanghai Institute of Ceramics, Chinese Academy of Sciences, Shanghai 200050, China

[4]University of Chinese Academy of Sciences, Beijing 100039, China

[5]Department of Applied Physics, Graduate School of Engineering, Osaka University, 2-1 Yamadaoka, Suita, Osaka 565-0871, Japan

[#]These two authors contributed equally to the work.

[*]To whom correspondence should be addressed: email: zhihaicheng@ruc.edu.cn, email: schen@ruc.edu.cn

**Abstract:** Nanocontact properties of two-dimensional (2D) materials are closely dependent on their unique nanomechanical systems, such as the number of atomic layers and the supporting substrate. Here, we report a direct observation of toplayer-dependent crystallographic orientation imaging of 2D materials with the transverse shear microscopy (TSM). Three typical nanomechanical systems, $MoS_2$ on the amorphous $SiO_2$/Si, graphene on the amorphous $SiO_2$/Si, and $MoS_2$ on the crystallized $Al_2O_3$, have been investigated in detail. This experimental observation reveals that puckering behaviour mainly occurs on the top layer of 2D materials, which is attributed to its direct contact adhesion with the AFM tip. Furthermore, the result of crystallographic orientation imaging of $MoS_2$/$SiO_2$/Si and $MoS_2$/$Al_2O_3$ indicated that the underlying crystalline substrates almost do not contribute to the puckering effect of 2D materials. Our work directly revealed the top layer dependent puckering properties of 2D material, and demonstrate the general applications of TSM in the bilayer 2D systems.

Following the discovery of graphene prepared by the micromechanical exfoliation technique,[1] much research intension has been paid on the two-dimensional (2D)[2-7] material due to its superior characteristic properties, including excellent physical,[8, 9] mechanical,[10, 11] chemical,[12, 13] electronic and optoelectronic properties.[14-17] One of the intriguing advantages of the 2D layered material is that their isotropic properties can be converted to be anisotropic, which results from the breaking symmetry owing to their intrinsic puckering behaviors. The puckering effect of 2D materials is due to the out-of-plane elastic deformation and the bending stiffness, and have attracted wide research interest.[18-19] The anisotropies of 2D material exerted powerful influences on the ultrathin flexible nano-electromechanical and thermal devices. [20]

Progress in the general understanding of isotropic to anisotropic relationships in 2D ultra-thin films requires experimental tools capable of imaging the details. Nanomechanical deformation has been extensively investigated by the atomic force microscopy (AFM) indentation experiments on suspended films.[21-23] Based on AFM platform,[24-27] friction force microscopy (FFM) and transverse shear microscopy (TSM), as the derivative mode of contact AFM,[28-32] now have been widely used to study the elastic deformation,[19, 31] crystal structure,[28] surficial ripples of 2D materials[8,34, 35] and so on. Such studies have revealed an important phenomenon in which nanomechanical contact behavior could be dependent on the number of layers and substrates. While this layer-dependent transient phenomena in 2D materials on different supporting substrates has not been experimentally investigated.

Here, we focused experimental investigation on the surface layer-dependent puckering behavior of 2D materials through the extensive TSM characterization for their crystallographic orientation imaging. Firstly, we directly verify that the puckering effect mainly occurs on the top layer of 2D systems through the exclusive crystallographic orientations of the top layer of $MoS_2$ on the $SiO_2/Si$ substrate. Secondly, the growth position of the adlayer graphene was definitely determined below the first layer during the CVD growth process by a combination of TSM and scanning Kelvin probe microscopy (SKPM) measurements. Finally, the TSM method was further generalized to realize the crystallographic orientation imaging of $MoS_2$ monolayer grown on the crystallized $Al_2O_3$ substrate.

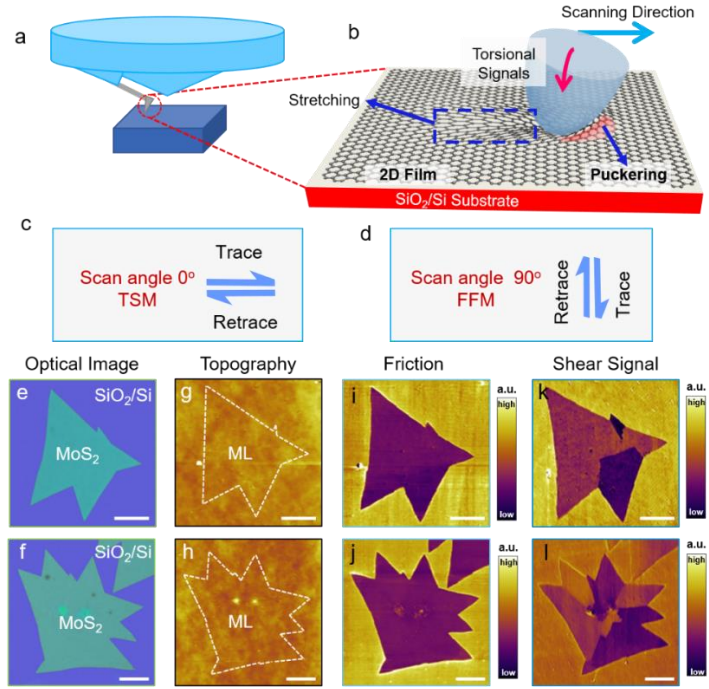

**Figure 1: FFM and TSM measurements of MoS$_2$ monolayer flakes on the SiO$_2$/Si substrate.** (a-d) Schematic diagram illustrating the working mechanism of TSM and FFM on the thin 2D materials. The mechanical model illustrates the puckering behavior of thin-film deformed by the loading and lateral force of the moving AFM tip. The scanning direction is perpendicular (parallel) to the cantilever axis in FFM (TSM). (e-h) Optical (e,f) and topography (g,h) images of two MoS$_2$ flakes without any crystalline grain contrast. The corresponding (i,j) friction (FFM) and (k,l) shear (TSM) signal images. Several grains in MoS$_2$ flakes are clearly resolved in the TSM images, but not in the FFM images. The scale bars are all 6 μm.

All of the AFM measurements were conducted using an Asylum research AFM system (MFP-3D infinity). Both FFM and TSM are the derivative modes of contact AFM mode, as illustrated in Fig. 1. For the 2D thin films, during the contact AFM scanning, the film shows puckering effect around the moving AFM tip, as shown in Fig. 1b. The puckered geometry induces the film to vertical relax at the front edge of the moving tip and simultaneously stretch the rear region of the moving tip. For FFM, the scan direction of the AFM tip is perpendicular to the long axis of the AFM cantilever, and then the friction property of the sample was obtained by detecting the torsion signal of the cantilever. For TSM, the scan direction of the AFM tip is parallel to the long axis of the cantilever, and then the shear property of the sample was obtained by detecting the torsion signal of the cantilever. For the 2D materials, the FFM measurements have shown that the friction force increases with the decreased thin film thickness due to the enhanced puckering effect. This puckering effect also enabled the TSM method to successfully image the orientation of the crystalline domains in star-shape MoS$_2$ monolayer flakes in our previously reported work.[27] While until now, it is still not clear about the toplayer-dependent puckering effect, and more experimental work should be performed to deeply understand this unique and typical nanomechanical behavior of 2D materials.

Firstly, the atomically thin CVD grown MoS$_2$ monolayer flakes were characterized by FFM and TSM imaging, as shown in Fig. 1. No crystalline differentiated contrast was observed in both optical and AFM topography images of MoS$_2$ monolayer flakes on the SiO$_2$/Si substrate.[7] Strain can induce the transformation from isotropy to anisotropy of both in-plane normal tension stress and shear stiffness, but shear stiffness exhibits much greater anisotropy.[31] Here, even enhanced by the puckering effect (strain applied by probe), the FFM signals still do not reveal any significant contrast dependence on the crystallographic orientations on the grains due to their almost isotropic stress-strain mechanical behavior. While the clearly crystalline grains were clearly resolved in the TSM images through the anisotropic shear-strain mechanical behavior. The reason for the difference of imaging contrast between FFM and TSM were discussed in our previous work.[31]

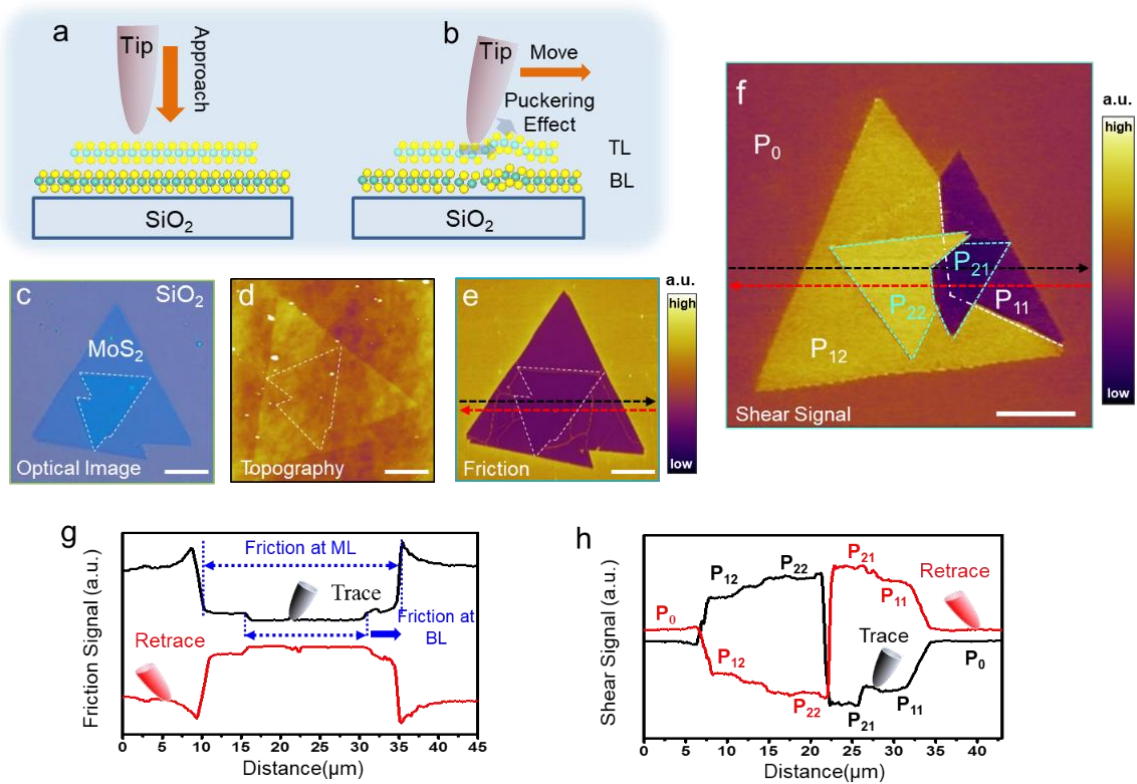

**Figure 2. Layer-dependent puckering effect of bilayer MoS$_2$ flake *via* FFM and TSM.** (a, b) Schematic model illustrating the puckering effect mostly occurred on the top layer (TL) of MoS$_2$ rather than bottom layer

(BL). (c-f) The optical (c), AFM topography (d), FFM (e) and TSM (f) images of bilayer MoS$_2$ on the SiO$_2$/Si substrate. The different MoS$_2$ grains are marked as P$_{11}$ and P$_{12}$ for the bottom monolayer, P$_{21}$ and P$_{22}$ for the top layer of bilayer MoS$_2$. The amorphous SiO$_2$/Si substrate is marked as P$_0$. (g, h) The line profiles of friction (g) and shear (h) signals in the trace and retrace directions marked by black and red line in (e) and (f). The scale bars are all 8μm.

In order to investigate the toplayer-dependent puckering effect, we further perform the systematic AFM characterization on the bilayer MoS$_2$ flake made of two bottom layer grains and two top layer grains, as shown in Fig. 2. Due to the van der Walls (vdW) interactions between the top and bottom layer, the adhesion and puckering effect will most occur between the AFM tip and the top layer when the AFM tip approach, contact and move on the top layer (Fig. 2a and 2b). Fig. 2c and 2d show the optical and AFM topography images of the bilayer MoS$_2$ flakes. The bilayer areas can be clearly resolved and were marked by the white dashed lines.

Fig. 2e shows the FFM images of this bilayer MoS$_2$ flake. No crystalline differentiated contrast was observed on both top and bottom layer grains. While it is noted that the friction signal on the bilayer area is slightly lower on the monolayer area, as shown in Fig. 2g, which is in consistent with the previously reports in 2D materials.[30, 38-41] Fig. 2f shows the TSM images of this bilayer MoS$_2$ flake. Both the top layer grains (P$_{21}$ and P$_{22}$) and the bottom layer grains (P$_{11}$ and P$_{12}$) were clearly resolved in the TSM image. The P$_{21}$ and P$_{11}$ grains are in a near-pristine 2H stacking. The top layer grain of P$_{22}$ shows almost the identical shear signal with the bottom layer grain of P$_{12}$ due to their pristine 2H stacking with the same but inversion-asymmetry crystallographic orientation. Since the vdW interaction between the top and the bottom MoS$_2$ layer is less than the bonding between bottom MoS$_2$ layer and substrate, the difference between shear signals (the trace and retrace) of top layer is larger than bottom layer (Fig. 2h). It is noted that the top layer grain of P$_{21}$ (and P$_{22}$) is stacked over both the bottom layer grains of P$_{11}$ (P$_{12}$), while no shear contrast was observed within the grain of P$_{21}$ and P$_{22}$. This result directly demonstrates that the puckering effect mostly happened on the top layer, then the corresponding puckering-induced shear signal in the bottom grains should be very small in contrast to the top layer grain.

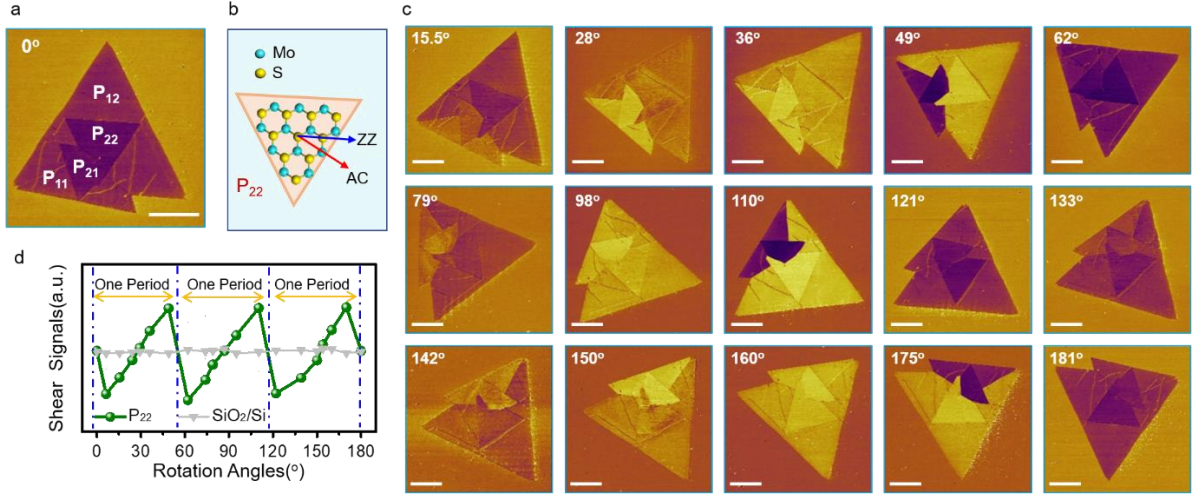

**Figure 3. Crystallography orientation dependency of the MoS$_2$ bilayer flake.** (a-c) Crystallographic labels for the AC and ZZ orientations of MoS$_2$. Every grain is given an identifier of P$_{11}$ and P$_{12}$ for the bottom monolayer, P$_{21}$ and P$_{22}$ for the top layer of bilayer MoS$_2$. TSM images obtained by rotating the sample in clockwise direction. All the images use the same color contrast. The scale bars are all 8μm. (d) Shear signal *vs* rotation angle curves obtained by randomly selecting P$_{22}$ domains with the underlying amorphous SiO$_2$/Si substrate.

It is necessary to further investigate whether the anisotropic domains of the top layer grains depend on the crystallographic orientation in the TSM images. In Fig. 3a and 3b, the crystallographic orientation (armchair and zigzag) of evert grain could be determined according to their primitive growth zigzag edges. By systematic TSM characterization with a series of clockwise rotations (Fig. 3c), the periods of 60° of the shear signals with the rotation angle are presented (Fig. 3d), which is consistent with the monolayer MoS$_2$ in our previous work.[31] There is an interesting phenomenon that a singularity (a jump in value) exists in the ZZ orientation [~60°, 120° and 180°]. This phenomenon can be explained by shear stiffness anisotropy along ZZ orientation and AC orientation. Under the same transverse deformation, a greater restoring force results in increasing lateral torsion of the cantilever near the ZZ direction. The amorphous SiO$_2$/Si does not show any obvious dependency on the rotation angle. These results further confirm our ideas on the toplayer-pendent crystallographic orientation imaging, also demonstrate the general application of TSM to determine the crystallographic orientations of 2D materials, especially in the twisted bilayer or few-layer systems.

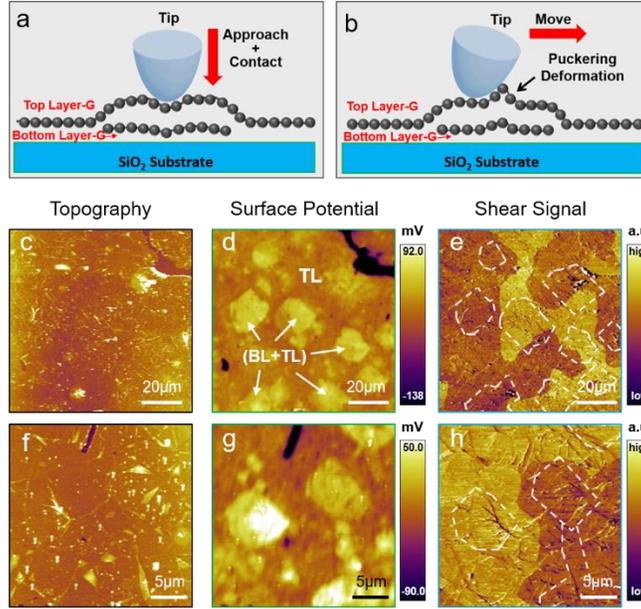

**Figure 4. AFM characterization of the CVD-grown bilayer graphene film transferred on the SiO$_2$/Si substrate.** (a, b) The schematic diagram of the bilayer graphene film with the second layer grown below the first layer. The puckering effect mostly happened on the top layer (TL) of graphene rather than bottom layer (BL). (c-h) AFM topography (c,f), SKPM surface potential (d,g) and TSM shear signal (e,h) images of the bilayer graphene transferred on the SiO$_2$/Si substrate. Both scratches and wrinkles are observed in the topography images due to the transfer process, and the bilayer graphene area could not be determined by the topography images. The bilayer graphene area is determined by its higher surface potential than the monolayer. The grains of top layer graphene are clearly determined in the TSM images. The bottom second layers are marked by white dashed lines in (e) and (f).

The toplayer-dependent TSM characterization can also be used to determine the special growth mode of bilayer graphene in combination with the SKPM measurements, as shown in Fig. 4. Fig. 4a and 4b schematically show that the adhesion/puckering effect between the static /moving AFM tip and 2D layers are mostly emerge in the contacted top layer graphene. Fig. 4c and 4f shows the AFM topography images of the bilayer graphene films transferred on the SiO$_2$/Si substrates (See details in supporting information). It is clearly that the bilayer graphene regions were not resolved in the topography images. Fig. 4d and 4g show the corresponding surface potential of the bilayer graphene films obtained by the SKPM measurements. The bilayer and monolayer graphene regions were clearly distinguished by their different surface potentials and in good agreement with previously reported work.[36] To further determine whether the second growth layer is underlying the first layer or not, we perform the TSM characterization of the graphene films, as shown in Fig. 4e and 4h. The crystalline domains of the top layer graphene were clearly observed, while no bilayer regions marked by the white dashed lines were resolved. We can directly conclude that the second graphene layer is underlying the first layer, confirm the underlying growth model of bilayer graphene.[36]

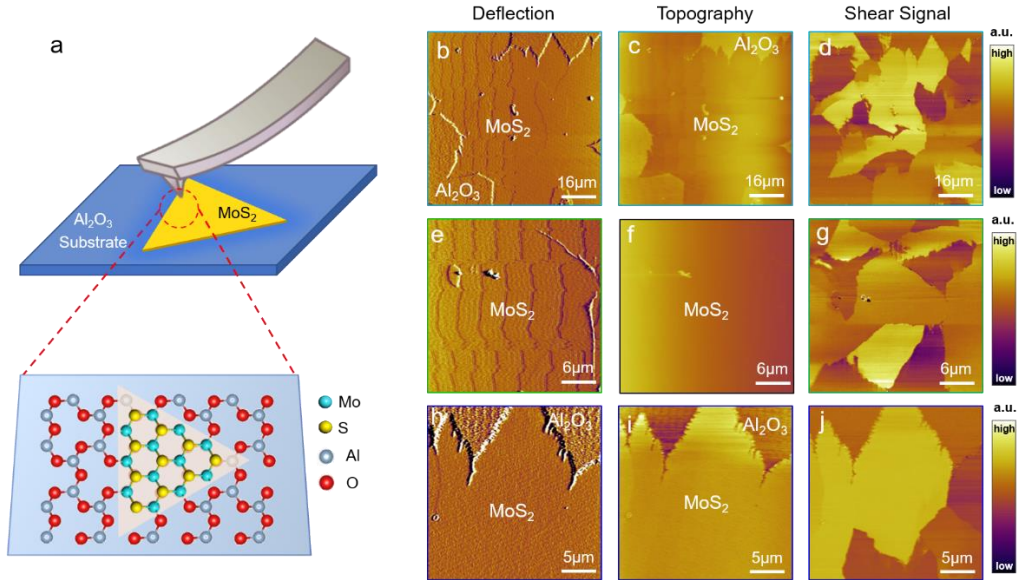

**Figure 5. AFM characterization of the CVD-grown MoS$_2$ monolayer on the Al$_2$O$_3$ substrate.** (a) Schematic showing the AFM imaging of MoS$_2$ monolayer on the atomically flat crystalline Al$_2$O$_3$ substrate. (b-j) The AFM deflection (b,e,h), topography (c,f,j) and TSM shear signal (d,g,j) images of the MoS$_2$ monolayer on the Al$_2$O$_3$ substrate. Many terraces with step edges of the crystalline substrate were clearly observed in the deflection images. The MoS$_2$ grains clearly visualized in the TSM images.

The previous TSM measurements of 2D materials were performed on the amorphous SiO$_2$/Si substrate. We found that amorphous substrate does not contribute any shear signals during the TSM imaging. While for the 2D materials on the crystalline substrates, whether the TSM method is still applicable or not, should be further investigated. The MoS$_2$ monolayer films directly grown on the crystalline Al$_2$O$_3$ substrate were further characterized by the TSM measurements, as shown in Fig. 5. The substrate terraces with straight edges were resolved in the AFM deflection images, indicating the crystalline nature of the Al$_2$O$_3$ substrate. No crystalline MoS$_2$ domains were resolved in the AFM topography and deflection images, but in the TSM images. Then we can conclude that the underlying crystalline substrate does not apparently contribute to the TSM shear signal comparing to the 2D layers with the puckering effect, and the TSM characterization can be generalized to 2D materials on the crystalline substrates.

In summary, we have identified the toplayer-dependent crystallographic orientation imaging of 2D materials, which enable the crystallographic orientation imaging of bilayer films with transverse shear microscopy. We experimentally demonstrate that only the top layer $MoS_2$ was crystallographic orientation imaged with TSM, which will be useful in the twist bilayer systems. The unique intercalation growth mode of the second layer graphene during the CVD growth process was confirmed with the toplayer-dependent TSM characterization and SKPM measurements. The crystallographic orientation imaging of $MoS_2$ monolayer on the crystalline $Al_2O_3$ substrate was further realized and demonstrated the general applications of the TSM method. Our research will be beneficial in understanding the nanomechanical behaviors of 2D systems and provides a convenient and powerful approach to facilitate the nondestructive crystallographic orientation characterization of various 2D atomic crystal systems.

The data that supports the findings of this study are available within the article and its supplementary material.

This work was supported by the National Natural Science Foundation of China (NSFC; grant numbers: 21622304, 61674045, 11604063), Ministry of Science and Technology (MOST) of China (grant number: 2016YFA0200700), Strategic Priority Research Program, Key Research Program of Frontier Sciences and Instrument Developing Project of Chinese Academy of Sciences (CAS; grant numbers: XDB30000000, QYZDB-SSW-SYS031, YZ201418), Grant-in-Aid for Scientific Research from Japan Society for the Promotion of Science (JSPS) from the Ministry of Education, Culture, Sports, Science, and Technology of Japan (JP16H06327, JP16H06504, JP17H01061, JP17H010610), and Osaka University's International Joint Research Promotion Program (J171013014, J171013007, J181013006, Ja19990011). Z. H. Cheng was supported by Distinguished Technical Talents Project and Youth Innovation Promotion Association CAS, the Fundamental Research Funds for the Central Universities, and the Research Funds of Renmin University of China (grant number: 18XNLG01). S. Chen appreciate the support from Beijing Natural Science Foundation (grant number: 2192024).

# Supporting Information for

# Toplayer-Dependent Crystallographic Orientation Imaging in the Bilayer Two-Dimensional Materials with Transverse Shear Microscopy

Sabir Hussain,[1,2,4, #] Rui Xu,[1, #] Kunqi Xu,[3] Le Lei,[1] Shuya Xing,[1] Jianfeng Guo,[1] Haoyu Dong,[1] Adeel Liaqat,[2,4] Rashid Iqbal,[2,4] Muhammad Ahsan Iqbal,[2,4] Shangzhi Gu,[1] Feiyue Cao,[1] Yan Jun Li,[5] Yasuhiro Sugawara,[5] Fei Pang,[1] Wei Ji,[1] Liming Xie,[2,4] Shanshan Chen,[1,*] Zhihai Cheng[1,*]

[1]Beijing Key Laboratory of Optoelectronic Functional Materials & Micro-nano Devices, Department of Physics, Renmin University of China, Beijing 100872, China

[2]CAS Key Laboratory of Standardization and Measurement for Nanotechnology, CAS Center for Excellence in Nanoscience, National Center for Nanoscience and Technology, Beijing 100190, China

[3]Key Laboratory of Inorganic Functional Materials and Devices, Shanghai Institute of Ceramics, Chinese Academy of Sciences, Shanghai 200050, China

[4]University of Chinese Academy of Sciences, Beijing 100039, China

[5]Department of Applied Physics, Graduate School of Engineering, Osaka University, 2-1 Yamadaoka, Suita, Osaka 565-0871, Japan

[#]These two authors contributed equally to the work.

[*]Corresponding authors: zhihaicheng@ruc.edu.cn    schen@ruc.edu.cn

**Materials and Methods**

**Chemical vapor deposition (CVD)-grown MoS$_2$ on amorphous SiO$_2$/Si substrate:** The MoS$_2$ layers were grown on 200nm SiO$_2$/Si supporting substrate by CVD method. MoO$_3$ (99.5% purity) and Sulfur (99.5% purity) were used as the precursor and reactant, respectively. MoO$_3$ powder (25mg) was placed in a quartz boat at the center of a furnace. The SiO$_2$/Si substrate (2×2 cm$^2$) was carefully place face down above the MoO$_3$ powder. The sulfur powder was heated to 180°C and was carried through Ar gas flow at 500sccm. The experiment was conducted at a reaction temperature ~750°C.

**Chemical vapor deposition (CVD)-grown MoS$_2$ on atomically flat crystalline Al$_2$O$_3$ substrate:** Put MoO$_3$ and S$_2$ powder in the designed powder. The MoO$_3$ powder is placed at the heating center, and the S powder is placed at a position about 17cm upstream from the MoO$_3$ powder. The atomically flat crystalline (Al$_2$O$_3$) supporting substrate is placed about 0.5cm above the MoO$_3$ powder, with polished side down. There is large amount of S powder, which can probably can cover the entire the bottom of the porcelain boat. The MoO$_3$ powder only needs a small particle visible to the nacked eye (0.01mg level). After placing the porcelain boat, immediately encapsulate the same gas, and then turn on the powder for heating. Heating conditions: heating to 870°C for 25-30 minutes from room temperature, then constant temperature at 870°C for 5 minutes, then natural cooling to room temperature. Gas: A mixture of 90% Ar and 10% H$_2$ has a flow rate of 35sccm.

**CVD-grown bilayer Graphene on amorphous SiO$_2$/Si substrate:** A Low-pressure CVD system was used for the growth of bilayer graphene on flat Cu foil. Typically, the Cu foil (Alfa Aesar, 25μm, 99.8%) was electropolished to smooth the surface as well as to remove the coating layer typically applied by the manufacturer. Then the pre-treated Cu foil was loaded into the reaction chamber of the quartz tube furnace. The CVD system was evacuated and heated up to 1030°C in 20sccm H$_2$ and held for another 30min to anneal the copper foil. After annealing, the growth was carried out with a 1.5 sccm methane flow for 1min followed by a 1sccm methane flow for another 20min. The hydrogen flow was maintained at 20sccm during the growth and the cooling process. After growth, the graphene film was transferred onto the amorphous SiO$_2$/Si substrate through the typical Polymethyl methacrylate (PMMA) assisted wet etching process.[4]

**Optical microscopy:** Optical characterization of MoS$_2$ layers was conducted by optical microscopy (ECLIPSE LV150N, Nikon).

**Atomic force microscopy (AFM) measurements:** All the AFM measurements were performed under ambient condition (MFP-3D Infinity, Asylum Research). The following functional AFM modes were used in this study:
**Transverse shear microscopy (TSM) and friction force microscopy (FFM)**: TSM and FFM are the derivatives of contact mode of AFM, which is optimized to measure the torsion signal between the tip and sample surface. For FFM, the scan direction of the AFM tip is perpendicular to the long axis of the AFM cantilever, and then the friction property of the sample was obtained by detecting the torsion signal of the cantilever. For TSM, the scan direction of the AFM tip is parallel to the long axis of the cantilever, and then the shear property of the sample was obtained by detecting the torsion signal of the cantilever. Both strain applied by substrate and induced by probe will enhance the TSM signal. All the FFM and TSM measurements were performed with a sharp Silicon AFM probe (AC160, Asylum Research) with spring constant of 22.2N/m was used over the course of investigation, with the scanning speed 1.5m/s.

**Kelvin probe force microscopy (KPFM):** KPFM measurements were carried out at room temperature. Commercially available conductive tip (conductive Pt/Ir coated AFM tip). KPFM measurements were recorded simultaneously AFM images with using the standard two-pass techniques: the first pass used to record the topographic images. Meanwhile second pass used to record the surface potential mappings by keeping the tip at selected the tip height with respect to recorded the topographic image. In our study the electrical tip was kept at a left height at 50nm to avoid the topographic artifacts.

## S1. Anisotropic response in merging flakes

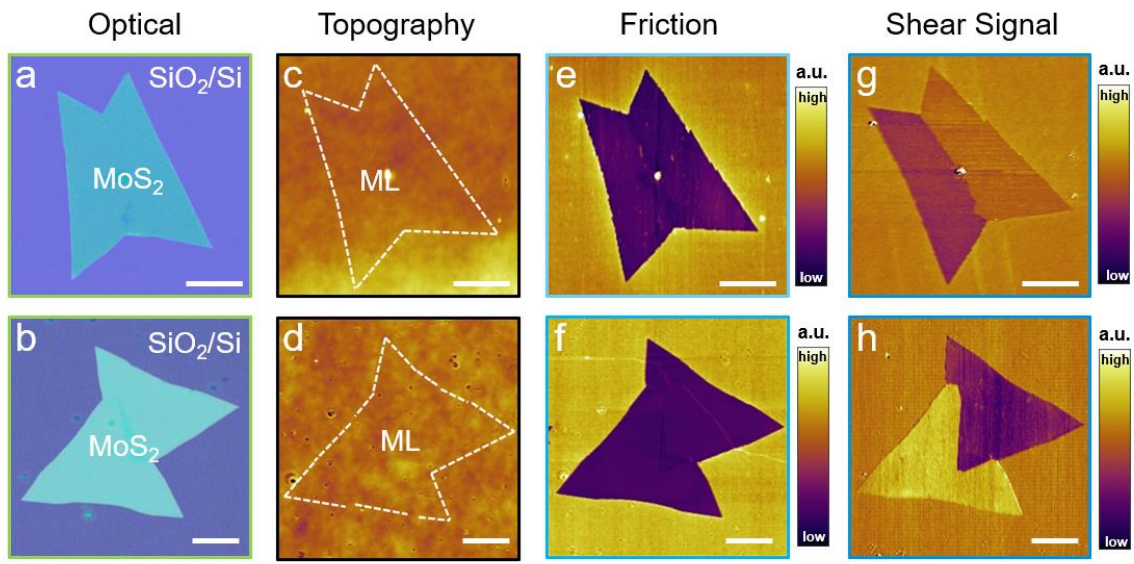

**Fig. S1.** (a, b) Optical, (c, d) topographic, (e, f) FFM and (g, h) TSM measurement of two typical MoS$_2$ merging flakes. Their different crystallographic orientations were clearly determined by the shear signal of TSM measurements.[1] The scale bars are all 5μm.

## S2. Loading force dependent imaging by TSM and FFM

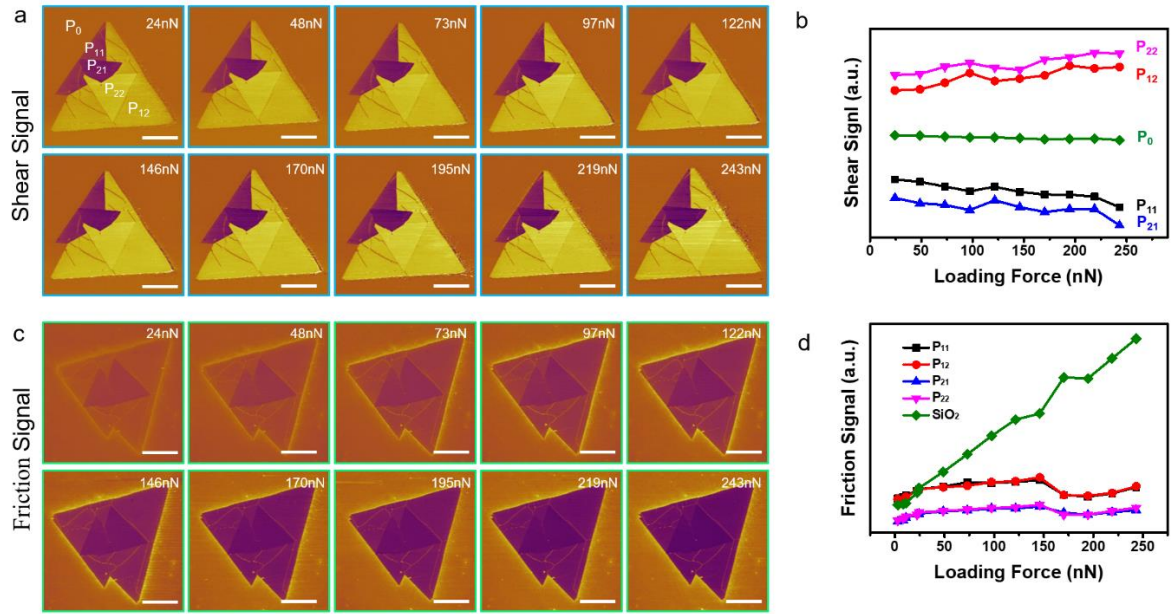

**Fig. S2. The *in-situ* TSM and FFM Characterizations with the increasing loading forces of AFM tip.** (a) TSM images under different loading force. (b) The profile lines illustrate shear signals vs loading force. (c, d) FFM images under different loading force and corresponding profile lines. The scale bars are all 8μm.

We employed the load dependent TSM and FFM techniques to characterize the bilayer $MoS_2/SiO_2$. Fig. S2a and S2b illustrate the nanomechanical contact response's signals and friction signals corresponding their loading forces. It can be seen that in profile line S2b and S2d the load dependent force can regulate the out of plan elastic deformation and friction response. Since it increases the contact area between the AFM tip and the sample surface, create a more puckered geometry and friction. Thus, the higher the loading force cause the higher stretch deformation and higher friction. During the low loading force, the contrast between the monolayer and bilayer is minimal, but at as the loading force increased cause the contrast grows as shown in figure S2(b). Fig. S2b illustrate the stretch deformation, where the brighter contrast has higher stretch deformation.[2, 3] According to our earlier work, the friction force is nearly proportional to the magnitude of the deformation.[1] For FFM, Fig. S2d illustrate the friction profile according to the load dependent by AFM tip from initiation to saturation point.

**S3. TSM images of multi-grain graphene films directly on the growth Copper substrate**

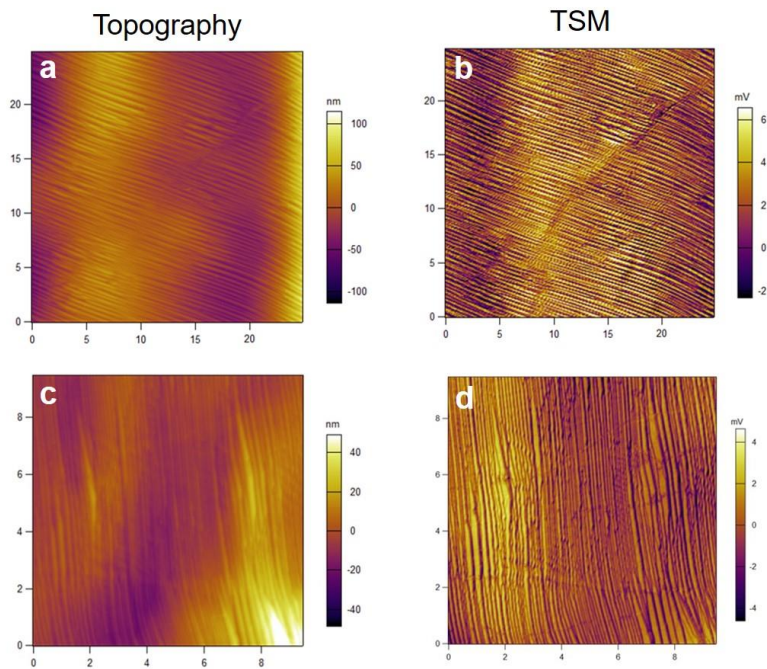

**Fig. S3.** (a, c) AFM topography and (b, d) TSM images of multi-grain graphene films directly on the growth Copper substrate. Different from the transfer graphene on the SiO$_2$/Si substrate, no grains were observed within the films, which should be due to the strong bonding between the 2D samples and substrates.